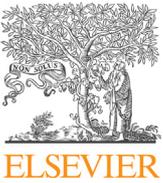
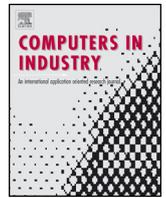

# Future developments in cyber risk assessment for the internet of things

Petar Radanliev[a], David Charles De Roure[a], Razvan Nicolescu[b], Michael Huth[b], Rafael Mantilla Montalvo[c], Stacy Cannady[c], Peter Burnap[d]

[a] Oxford e-Research Centre, Department of Engineering Sciences, University of Oxford, UK
[b] Imperial College London, UK
[c] Cisco Research Centre, Research Triangle Park, USA
[d] School of Computer Science and Informatics, Cardiff University, United Kingdom

A B S T R A C T

This article is focused on the economic impact assessment of Internet of Things (IoT) and its associated cyber risks vectors and vertices – a reinterpretation of IoT verticals. We adapt to IoT both the Cyber Value at Risk model, a well-established model for measuring the maximum possible loss over a given time period, and the MicroMort model, a widely used model for predicting uncertainty through units of mortality risk. The resulting new *IoT MicroMort* for calculating IoT risk is tested and validated with real data from the BullGuard's IoT Scanner (over 310,000 scans) and the Garner report on IoT connected devices. Two calculations are developed, the current state of IoT cyber risk and the future forecasts of IoT cyber risk. Our work therefore advances the efforts of integrating cyber risk impact assessments and offer a better understanding of economic impact assessment for IoT cyber risk.

© 2018 The Authors. Published by Elsevier B.V. This is an open access article under the CC BY license (http://creativecommons.org/licenses/by/4.0/).

## 1. Introduction

The Internet of Things (IoT) is still in its infancy: all actors and products in this space are early adopter stage. At present, we therefore need to understand the economic impact related to IoT and to develop reusable models where knowledge and findings can be captured, transferred, and used to create substantial effect. This is because the corporate adopters like financial markets and banks are entering in areas of new and potentially disruptive technologies such as IoT, blockchain, Artificial Intelligence. Corporate adopters can benefit from such new models as they can point out any metaphorical landmines and pit traps whose avoidance can positively influence the present IoT market. It is likely that corporate adopters are rushing into IoT without clear ideas of what the possible impact of failure could be, often in pursuit of apparent financial advantage. Hence, this article proposes a new assessment model for articulating possible impacts and calculating the economic impact of IoT cyber risk.

The IoT represents the idea of networked objects communicating their data among themselves. This communication can be controlled across other objects, systems and servers. Communication between objects, networks and humans involves conscious and/or unconscious actions from one IoT agent or device directed to one or several IoT agents or devices. The term 'IoT' was created in 1999 [1] and the first IoT principles were published shortly after in the book 'When Things Start to Think' [2]. According to Gartner's IT Hype Cycle, the IoT market adoption should take 5–10 years, as of 2012 [3]. The increased adoption of the IoT includes interoperability across multiple categories of cyber-physical systems [4], integrating technologies related to smart grids, smart homes, intelligent transportation, manufacturing and supply chain and smart cities. Such new technologies come with new types of risk that existing risk assessment/management methods are neither designed to anticipate nor predict.

Safeguarding an IoT deployment, while simultaneously harnessing its economic value, requires systematic consideration of multiple risk factors. Cyber-attacks are occurring more frequently and increasingly target IoT devices (for example the Mirai botnet). With a constant growth of both the attack surfaces and attack capabilities, the severity of future attacks on IoT/IT systems could be much greater than what has been observed to date. To predict the severity of future attacks, this article adapts two established models for predicting uncertainty - the Cyber Value at Risk [5] model and the MicroMort [6], for calculating the economic impact of IoT cyber risks.

A critical question for government policy as well as for private sector business strategies regarding IoT connected products, platforms and services is the sufficiency of cyber security measures

*E-mail addresses:* petar.radanliev@oerc.ox.ac.uk (P. Radanliev),
david.deroure@oerc.ox.ac.uk (D.C. De Roure), r.nicolescu@imperial.ac.uk
(R. Nicolescu), m.huth@imperial.ac.uk (M. Huth), montalvo@cisco.com
(R.M. Montalvo), scannady@cisco.com (S. Cannady), p.burnap@cs.cardiff.ac.uk
(P. Burnap).

and methods to minimise cyber risk that accompanies IoT deployments. The answer must be partially addressed by economic analysis, such as cost and frequency analysis of cyber-attacks [7]. Such analyses would complement and inform better the creation and operationalization of frameworks and methodologies for mitigating the economic impact of cyber risk of commercial use of IoT connected products and services [8]. This paper proposes such a new economic model for cost and frequency analysis of IoT cyber-attacks for IoT deployments.

Effective mitigation of the negative economic impact of IoT cyber risk should make use of statistical techniques that measure and quantify the level of risk in a given timeframe. The model Value at Risk (VaR) [9], e.g., measures the maximum possible loss during a specific time period. The Value at Risk has been advanced recently into the Cyber Value at Risk (CyVaR) model, specifically in order to address the specific problem of financial loss from cyber risk [10].

The research methodology used in this article adopts the Cyber Value at Risk (CyVaR) and the MicroMort (MM) approach and evaluates other cyber risk assessment approaches in order to build a new model for calculating the economic impact of IoT cyber risk. There is limited research on the economic impact of cyber risk. There is even less research on the economic impact related to cyber risks within different IoT verticals. By IoT verticals, usually we refer to IoT applied in conventional vertical markets such as health care, manufacturing, building automation, finance and insurance, logistics, and retail. Currently, the economic impact of IoT related cyber risks is assessed by applying methodologies established before the development of IoT verticals [11], which is likely to overlook IoT-specific aspects. It is challenging to devise a methodology that would be effective in all IoT verticals, given their number and diverse nature, and doing so would blur the interpretation of IoT is its usage. Also, IoT verticals may offer consumer/domestic applications and industrial IoT without a clear appreciation that the vulnerabilities, risks and impact of failure are very different for these domains. Therefore we cannot expect a 'one size fits all' approach to evaluating the IoT cyber risks in IoT vertical markets. Rather, we reinterpret the meaning of IoT verticals as those aspects that can be seen in different verticals as most promising: automated, digital, social machines, cyber-physical and coupled systems. This specific understanding of the term IoT verticals, enables the re-categorisation of what is commonly referred to as verticals into '*vertices*', a concept that offers a more abstract classification of IoT verticals (e.g. digital) based upon the underpinning pillars of vertices (e.g. health care).

This re-categorisation is inspired by characterisation of devices [12] and was deemed necessary as present day infrastructure systems are far more complex than the conventional verticals, creating novel risks for failures. And risk in an IoT deployment might extend to many entities. For example, an interruption in services delivered by a digital vertical (e.g. automated manufacturing), creates impact to different vertices (e.g. smart grid or smart city) and would impact many different businesses, agencies and individuals. A new impact model and assessment methodology could thus anticipate economic impact of cyber risks and benefits it its mitigation for the IoT ecosystem.

## 2. Literature review

Our literature review probes the essence of impact measurements and sources of probabilistic data for cyber risk impact assessment methods. This investigation results in a taxonomic classification of impact measurement units into an intuitive scheme and a categorisation of IoT cyber risk vertices.

### 2.1. Cyber risk impact assessment

The IoT security standards, such as the PAS 754 *Software Trustworthiness Standard* and the forthcoming BS10754:2018, provide definitions for trustworthiness of information technology, cyber physical systems and present governance and management specifications [13]. IoT security standards also mean to enable trustworthy IoT systems [14,15] with minimal human intervention [16]. Such autonomous IoT, requires a risk-based adaptive assessment framework [17] for risk analysis [18,19]. Risk-based adaptive assessment involves the following abilities: to predict problems, to predict impact, to implement planned actions, to maintain focus on risk mitigation, and to reduce risk exposure [18]. Since cyber risk is constantly evolving, such risk has not been clearly quantified through historical measures [20]. A commonly quoted figure stated is a annual global loss of $1 trillion to cybercrime, but estimates range from: 300bn and $1tn [21], $400bn to over $575bn [20], or $400bn to over $2tn [22]. The differences in these figures suggest that they are rough estimates at best, and the real economic impact of cyber risk remains unknown [22]. Also, factors other than cybercrime may influence cyber risks: cyber losses may occur because of system failures, interdependencies and cascading risks, as well as through simple business continuity issues – e.g., an outage by a major cloud service provider. The main difficulties in calculating the economic impact of cyber risk are the lack of suitable data and the lack of a universal, standardised framework for assessing cyber risk [23]. There is furthermore the need to quantify accumulated risk on a shared technology platform (such as cloud computing on a shared platform) and the digital supply chain [24].

### 2.2. Cyber risk assessment in the Internet of Things domain

In this section, the literature review is organised in a taxonomic classification of cyber risk assessment requirements. The categorisation follows recommendations from literature [26] by categorising cyber risk assessment requirements into (1) risk identification assessment strategy; (2) risk estimation strategy; and (3) risk prioritisation strategy.

IoT capabilities create new types of cyber risk [26], which are neither anticipated nor considered in existing cyber risk assessment standards. To adapt the current cyber security models, firstly the specific IoT cyber risk vertices are identified in Table 1. By risk vertices, we refer to IoT vectors from particular attack approach used, to exploit IoT vulnerabilities. Subsequently, these specific risk vertices need to be integrated in a holistic cyber risk impact assessment model. Because integrating IoT technology in the communications networks of critical infrastructure implies major ethical aspects that humans should be able to be aware of and comprehend, while also benefiting from maximum possible levels of trust and privacy. Integrating IoT technology in the communications networks also triggers question on data ownership, data privacy and economic lifespan of digital assets, it has been established that digital assets can outlive humans [25], triggering the question of data ownership after the end of data owners' lives. Some studies have simplified the topic with the assumption of limited economic lifespans for all classes of digital assets [24].

An integration of IoT risk vertices into reliable cyber security frameworks would help with preventing abuse originating from malicious interventions, including those perpetrated by organised crime, terror organisations or state-sponsored aggressors. An analysis of the complete economic impact of data compromises would empower the communications network providers or data owners with the ability to create clear, rigorous, industry-accepted mechanisms to measure, control, analyse, distribute and manage critical data needed to develop, deploy and operate cost-effective cyber security for critical infrastructure.

## 3. Research methodology

This section outlines the methodology applied in the research reported here. The methodology began with a literature review to create a taxonomical categorisation of impact assessment classes. In Section 4, a SWOT analysis of existing frameworks discusses the complexities of designing a new impact assessment. Finally, in Section 5 a new quantitative model is developed, by adopting the Cyber Value-at-Risk (CyVaR) model [36] and MicroMort (MM) [6] for impact assessment of IoT cyber risk. The CyVaR model is based on the notion of *value at risk*, widely used in the financial services industry and based on a probabilistic approach to estimate the likely loss from cyberattacks over a given period. This model attempts to understand the economic impact of cyber risk for individual organisations [23] and has been promoted for standardisation of language, models and methods [37]. Building upon the CyVaR, a unifying economic framework proposed the use of measurement units for cyber risk [24]. Other cyber value analysis methods have proposed to calculate the cost of different cyber-attack types [38], but they lack data for validating their models. This lack of data has motivated the development of a proof-of-concept method [23], based on *data assumptions*. The weakness in the latter approach is that economic impact is calculated on organisations' *stand-alone* cyber risk, because credible data assumptions can only be made on intra-organizational data. However, business impact for the same risk can vary widely between companies based on the specific circumstances of each company [39]. Furthermore, that approach ignores the correlation effect of organisations sharing infrastructure and information, and by default, that of sharing cyber risk exposure. Cyber risk exists in multiple physical, information, cognitive, and social domains – to mention software, hardware, firmware, adjacent systems, energy supplies, supply chains – and the economic impact is related to these closely interconnected systems. This close interconnection of disparate systems increases the probability of 'cascading impacts' [20].

To summarize, the work reported in this article therefore applied multidisciplinary methodologies, along with established risk measurement methods such as MicroMort (MM) to define individual risk units and Value-at-Risk (VaR) for measuring market cyber risk.

## 4. Analysis of cyber risk frameworks, methods, systems and models

Existing risk assessment methods based on Return on Investment (ROI) and Net Present Value (NPV), include broad sets of criteria, e.g. 'economics of privacy' [40], 'optimal amount to invest' [41], 'risk averseness' [42]. However, cyber risk covers more elements than thos pertaining to the financial cost of information security. Thus, we need a method that can integrate cyber risks directly with economics [24]. Because the motivation for cyber risk can be other than purely financial (e.g. espionage) yet still create economic impact, said impact should be calculated in terms of averages and for the most severe scenarios [23].

To make such calculations for impact assessment reasonably precise and meaningful, different modelling approaches need to be integrated into a new and more reliable impact assessment model. Such integration is important for the reasons aforementioned. Insurance companies need to better understand such the integration of diverse approaches and how it may influence cyber risk assessments. This understanding is often furthered through a **S**trengths, **W**eaknesses, **O**pportunities and **T**hreats (SWOT) analysis. We now summaize the SWOT analysis of the reviewed frameworks, methods, systems and models.

The Operationally Critical Threat, Asset, and Vulnerability Evaluation (OCTAVE) [43] investigates recovery impact areas via a questionnaire. OCTAVE does not address the quantification of risk impact; more concerningly, it does not provide examples of how the methodology is applied. The greatest opportunities OCTAVE presents are applicable to small companies with limited resources, because it is free and easy to use. In terms of how it fits with other approaches, it could be used as a starting point in risk assessment. OCTAVE method is fairly complex to understand and this should be considered as a threat, considering that the main users for this method could be from small companies with limited resources.

The Threat Assessment & Remediation Analysis (TARA) [44] is focused on targeting only the most crucial exposures. This represents a different approach to other predictive frameworks focused on defending all vulnerabilities. TARA is similar to OCTAVE in terms of not focusing on the quantification problem. The opportunities TARA creates are mainly based on its complimentary nature. TARA, e.g., can be applied in combination with other approaches such as the second step after OCTAVE. The greatest strength of TARA also represents its greatest weakness: it only focuses on the greatest risk, other risks are ignored.

The Common Vulnerability Scoring System (CVSS) [45] applies simple mathematical approximations to translate experts opinions into a numerical score of vulnerability severity. CVSS is based on a small number of variables. Although the scoring is based on 0–10, the actual system contains only 3 color-coded levels. This effectively means that score simplifications result in different vulnerabilitis having the same level or similar score. The greatest opportunity for the CVSS is its potential to increase the number of colours for its color-coding system and by doing that, to enhance the visibility of different scores. The threat from CVSS is that the mathematical formalism is fairly basic, and the mathematical score may wrongly convey a sense of certainty and truthfulness.

**Table 1**
Taxonomic classification of cyber risk assessment requirements.

| Classification of categories | |
|---|---|
| Risk identification assessment strategy | This should cover: espionage, theft, or terrorist attacks, which in effect requires electronic and physical security [27,28], to anticipate and mitigate any risks [29]. |
| | Risk identification should be supported with forensics, prognostics, and recovery plans, for analysis of cyber-attacks and for coordination with agencies responsible to identify external cyber-attack vectors [20]. |
| Risk estimation strategy | This should cover information assurance, data security and protection for data in transit, from physical and electronic domains and storage facilities [20,30,31]. |
| | This also requires supply chain risk analysis of components introduced in the supply chain [32], modified from its original design to enable a disruption or an unauthorised function [20,33]. |
| Risk prioritisation strategy | This should limit the access of source code to crucial personnel and should provide software assurance and application security for eliminating deliberate flaws and vulnerabilities [27]. |
| | To prevent continuation of cyber-attacks, risk prioritisation should focus on information sharing and reporting. Fast cyber-attack reporting and shared database resources should also be developed [20,34,35]. |

The Exostar system (Exostar) [46] is focused on cybersecurity of supply chains and regulatory compliance of supply-chain partners. This also represents a different focus as most of the other approaches reviewed are focused on the companies' stand-alone assessments. Exostar, however, does not assess a company's stand-alone risk. Rather, its focus is entirely on supply-chain partners. The greatest opportunity for the Exostar system is to integrate with other approaches and to combine companies' own risk assessment with the supply chain risk assessment. The main threat from applying the Exostar system is that the results depend on the supply-chain partners providing answers in a questionnaire. If the answers are incomplete or incorrect, their use is likely to lead to incorrect results.

The Capability Maturity Model Integrated (CMMI) [47] is focused on enterprise risk and the risk in the product development lifecycle. This approach is complimenting Exostar and, in combination, these two approaches would seem to present a more comprehensive model. The weakness of the CMMI is that it only points out vulnerabilities, it does not provide guidance on addressing identified vulnerabilities. The greatest opportunity for the CMMI is the continuous updates with the ISO 9001. Since the two approaches are related, there are opportunities in updating the CMMI with the ISO 9001 criteria. The threat from applying the CMMI is that the weaknesses would be easy to identify, but difficult to correct.

The approach of the National Institute of Standards and Technology (NIST) in [48] is distinct from other approaches because it represents a collection of standards and guidelines. The NIST framework covers risk assessment and risk management, which are two separate but complimenting operations. The weakness of the NIST framework is that it is effectively a framework with documented processes, which could be made usable by the development of tools that automate such processes. The greatest opportunity for the NIST framework is to replace the extensive use of acronyms it offers with an automation tool. For example, since these abbreviations seem to be created in an Excel file, the use of pivot tables could enable companies to select one section of the criteria in the process of reviewing the company's compliance. The threat from applying the NIST framework stems from its large size, extensive scope, and lack of automation support; the latter makes the process of documenting the updated recommendations time-consuming. Other approaches could be updated faster, making made recommendations perhaps less effective due to the extra time and work required to reflect them in documents.

The Factor Analysis of Information Risk Institute (FAIR) [49] model is singled out by its use of a quantitative approach for impact assessment. The main strength of the FAIR approach is that it recommends acceptable levels of exposure. The opportunity for the FAIR is to develop a standardisation reference that would not be based on voluntary compliance and consensus. This would effectively address the greatest weakness of the ISO standardisation approach. The threat from the FAIR model is that it promotes a commercial software, which comes at a cost. This creates problems for standardisation of the risk assessment approach, as the standardisation could be perceived as a platform for promoting commercial software.

The International Organisation for Standardisation (ISO) [50] differentiates itself with its focus on global standardisation of risk assessment. While the ISO approach promotes compliance and standardisation, it is effectively based on voluntary compliance and the standardisation is based on consensus. Yet, one can hardly imagine how a consensus can be reached in some areas of risk assessment where different countries and partners would have conflicting interests. The weakness is therefore that if consensus cannot be reached, some areas may not be included. Even if consensus is reached, there is no mechanism to ensure compliance. The greatest opportunity for the ISO approach is the potential to expand into a global standardisation reference for cyber risk assessment. The greatest threat to ISO is that it depends on the voluntary compliance and consensus from 161 countries and 778 technical committees and subcommittees. The coordination and implementation of such complex structure, based on voluntary compliance and consensus, is difficult to advance with the same speed as the cyber risks themselves evolve.

The RiskLens [51] is a quantitative assessment software promoted by the FAIR model. The main criticism of this software is that it is not peer-reviewed and effectively represents a black box whose results you have to trust, without being able to understand the process of how the quantification results are calculated. The greatest opportunity for the RiskLens software is to subject its algorithms to a peer-reviewed process. While this could conflict with the commercial nature of the software, the assessment should not lose value from the peer-review. There are alternatives to protecting the commercial values of the software, other than keeping it as a black box. The threat is that with the current lack of peer-review validation, the validity and confidence in results could easily be questioned.

The Cyber Value at Risk (CyVaR) [10], similarly to the RiskLens software, represents a quantitative approach and can be based on Monte Carlo simulations. The CyVaR model is also promoted by the FAIR institute. The criticisms of this approach are too many to cover here. Some of the most concerning weaknesses include the lack of required data (e.g. standard deviation, mean and median of recorded losses from cyber-attacks, etc.) to conduct the assessment. The greatest opportunity for the CyVaR model is to integrate new types of risks, e.g. the IoT cyber risks. The threat to applying CyVaR is its complexity. Unless simplified in an automated approach, enterprises could find it difficult to understand and apply the approach.

The above SWOT analyses explain multiple issues in building one quantitative model that would control all complexities of cyber risk assessment. Existing cyber risk frameworks and methodologies are constrained by a number of limitations. For instance, cyber risk assessment frameworks are based on security control domains and assess security posture, but are not effective in assessing high-risk loss scenarios developed around critical digital assets [24]. Furthermore, cyber risk assessment methodologies have created an inconsistency in measuring cyber risk, because of the absence of a common point of reference [24]. It is argued that the common point of reference should be represented as a unified approach for cyber risk assessment [8,52].

There are additional complexities not discussed in the analysis, because they are almost impossible to quantify. For example, in information assets such as intellectual property of digital information, the future value is lost even under early detection [23]. Therefore, the economic value of digital assets has to reflect their economic functions before their value can be properly assigned [24]. In addition, analysing the economic impact of cyber risk is also complicated because of the impact on brand reputation, the cost of downtime, legal liability, cost of intellectual property loss, and many other variables. Media coverage of cyber risk alone has created such significant economic impact that managing risk has become 'imperative' [21]. The following section, develops the design of a holistic model for calculating the economic impact of IoT cyber risks.

## 5. The model

We need a reliable model for costing cybercrime [53] and the first step in developing a costing model for IoT cyber risk is to determine the cybercrime units of costings. To determine the risk of cybercrime, we refer to established methods for calculating risk.

**Risk = Likelihood × Consequences**, and cyber-risk can be defined as a function of:

R = {si, pi, xi}, i = 1, 2, ..., N,

R – risk; s – the description of a scenario (undesirable event); p – the probability of a scenario; x – the measure of consequences or damage caused by a scenario; N – the number of possible scenarios that may cause damage to a system.

The described model is classical, but the question remains how p and x be estimated. Since we do not have the precise measurements and concrete number of the IoT cyber risks, an answer is difficult to present and justify with a degree of certainty that the estimation is correct. We can, however, discuss how p and x can be estimated if we had the required data. Looking at this problem from the perspective of probability theory, the cyber risk from one IoT vertical is represented as T and the cyber risk from its associated IoT vertices is represented as Y, and the Tx is the margin. Then the probabilities are defined as follows:

1. P(Tx|Y)=P(Tx&Y)/P(Y)P(Tx|Y)=P(Tx&Y)/P(Y)
2. P(Tx|T)=P(Tx&T)/P(T)P(Tx|T)=P(Tx&T)/P(T)
3. P(Tx)=P(Tx|Y)P(Y)+P(Tx|T)P(T)

In a simulated scenario where the T = 0.9 and the Y= 0.1

4. P(Y)+P(T) = 1

The P(Y) is the probability that the risk impact is related to the IoT vertices, and P(T) is the probability that risk impact is related to the IoT vertical, and P(T) is the marginal probability.

We can therefore compute these probabilities as:

P(Tx)=P(Tx|Y)(1−P(T))+P(Tx|T)P(T) =P(Tx|You)+P(T)[P(Tx|T)−P(Tx|Y)]

and therefore

P(T)=[P(Tx)−P(Tx|Y)]/[P(Tx|T)−P(Tx|Y)]

P(T)=[0.6 − 0.9]/[0.1 −0.9] = 0.375

Finally, let us consider the case in which the probability of Y is sequentially dependent on the probability of T. In other words, if the two probabilities are not independent of one another, and the accuracy of Y somehow depends on the accuracy of T. To give this scenario some context, let us assume that if the probability of T is defined in a state A, then Y is 0.3, but if the state of probability for T is B, then Y is 0.2, and if the state of probability for T is C, then Y is 0.1.

Then this would change our formula to:

P(Tx|T,C)=P(Tx&T|C)/P(T|C)

where P(C) is the probability of T being in a different state.

To build a model for calculating the impact of IoT cyber risk, we need to combine established risk models [24], such as MicroMort (MM) and Value-at-Risk (VaR) for measuring market risk and calculate new cyber risk units for IoT MicroMort (IoTMM) for these models. We then define IoT MicroMort2 (IoTMM2) as the value of reducing the risk by a given IoTMM.

The economic functions of IoT assets require an International IoT Asset Classification (IIoTAC), to enable calculating profits and losses from individual IoT assets (e.g. balance sheet, cash flow, etc). The term is chosen to be compliant with the proposed International Digital Asset Classification (IDAC) [24].

IoT digital assets can be categorised as: (1) IoT core value assets (IoTCA), digital assets that are directly part of goods or services that T profits from; (1a) IoT digitised assets (IoTDA), goods and services digitised from traditional goods and services; (1b) IoT assets born digital, representing things and services that are intrinsically digital; and (2) IoT operational assets (IoTOA), representing assets that support the creation, consumption and distribution of IoT goods and service.

A thing's ($T_h$) IoT composition can be described by the ratio of its core value assets to operational assets: IoTCA:IoTOA={ci,pi}:{oj,qj} i = 1,2, ..., Nc, j = 1,2, ...,No where

IoTCA – $T_h$'s core value assets; IoTOA – $T_h$'s operational assets; c – a type of asset listed in IDAC which is of core value to $T_h$; p – $T_h$'s core digital asset c; o – a type of asset listed in IDAC which is of operational value to $T_h$; q – $T_h$'s operational asset o; Nc – the number of core value assets in $T_h$; No – the number of operational assets in $T_h$.

By using the same formula, the IoTDA (digitised assets) to IoTAD (assets born digital) ratio of $T_h$ can also be calculated. The digital value composition of $T_h$ describes its nature of innovation, e.g. traditional goods have a high IoTOA:IoTCA ratio, while software has a high IoTCA:IoTOA ratio and a high IoTAD:IoTDA ratio. Other valuation parameters are: Intrinsic value of a IoT digital asset can be determined through fundamental analysis without reference to its market value. Market value of a IoT digital asset is the price at which the digital valuable would trade in a competitive market. Subjective value of a IoT digital asset is determined by the importance the enterprise places on $T_h$.

Following these valuation parameters, the value of IoTDA assets is directly converted from their physical equivalents. The value of IoTAD assets requires their own valuation analyses. IoTCA assets can be valued with Business Impact Analysis (BIA). Therefore, the formula of the existing economic theory of value to digital asset, the total digital value of $T_h$ can be calculated as:

$$V = \sum_{i=1}^{Nc} CVi + \sum_{j=1}^{No} OVj$$

where:

V – total digital value of $T_h$; cv – value of core value asset c of $T_h$; ov – value of operational asset o of $T_h$; Nc – the number of core value assets in $T_h$; No – the number of operational assets in $T_h$.

This valuation requires Key IoT Cyber Risk Factors (KIoTCRF) correlated with a $T_h$'s risk profile. Established Key Cyber Risk Factors (KCRF) risk categorisations [24] can be adapted to IoT, where technological factors are related to the usage of technology. Non-technological factors are related to people, process, socio-economic and geo-political factors. Inherent factors are related to $T_h$'s nature of business, industry, core operations, goods and services. Control factors represent $T_h$'s control effectiveness against cyber loss. Therefore, the $T_h$'s residual cyber risk can be calculated as:

**Residual cyber risk = inherent risk ÷ control effectiveness**.

This valuation allows for MM to be applied to define cyber risk units for both classes of D assets (IoTDA and IoTAD) and to define IoT MicroMortD (IoTMMD) for a given class D digital assets as 1 in a million probability of its digital death (where a $T_h$ would lose all economical value), where the value of 1 IoTMMD is the amount of money T is willing to pay to reduce 1 IoTMMD for its class D assets.

IoT residual risk IoTMM is not yet statistically available. However, when it will become statistically available for various types of IoT assets, it could be aggregated with asset values to generate a cyber VaR curve, representing T's residual cyber risk:

$$VaR = \sum_{i=1}^{n} Vi \oint Di$$

To compute the cyber VaR curve, historical simulation and Monte Carlo simulation can be used, where VaR is Value-at-Risk for all IoT digital assets of $T_h$; $T_h$'s digital asset inventory D = {D1, D2, ...,

Dn}; the value of each asset V = {V1, V2, . . ., Vn}; and Di is the amount of residual risk Di is exposed, measured in IoTMMD. Monte Carlo methods can generate a large number of paths using repeated random sampling to produce a probability distribution. In this scenario, the risk measure IoTMM2 (which is the value of reducing the risk by a given IoTMM) can be defined as a 12-month IoTMM2 VaR representing the loss limit $T_h$ can afford from cyber incidents. Where IoTMM2 is the cost $T_h$ is willing to pay to reduce its IoTMM2 risk by 1%. The VaR can be calculated for 12 months to represents cyber risk exposure over one financial year, as required for budget planning in ERM frameworks.

The proposed valuation depends on advanced data analytics and its ability to support a trajectory of exponential growth. We then have the advantage of storing and processing large datasets, hence the main obstacle is not the lack of capabilities to compute datasets, but to break down non-technological barriers and establish a wide range of data points in the proposed categories. It may take years or decades to validate the economic impact of IoT cyber risk, because of the time required for data collection. However, it is important to set the categories in order for the data collection to be performed in a structured manner.

### 5.1. Justification for the model

The proposed model is based upon the Cyber Value at Risk (CyVaR) [5,10] which represents adaptation of the traditional Value at Risk (VaR) model [9]. VaR is a statistical technique used to express the level of financial risk, or the financial risk of a given investment portfolio. The flexibility of application on a unit level or on an enterprise level, enabled the adaptation of the VaR to IoT risk vectors and vertices. Similarly, since the CyVaR is already adapted with probabilities for calculating the likely losses of cyber-risk over a given period of time, the adaptation of the CyVaR for IoT cyber risk simply represents following the trend in cyber risk assessment approaches. Instead of continuously building new models that would take decades to verify, the statistical techniques used in the new model are time tested and verified. The justification for using the MicroMort [54] was based on the lack of required data to apply the CyVaR statistical technique. The economic value of life has been researched even earlier [55] than the MicroMort. The MicroMort is used for many different calculations, starting from the risk of death on the day we are born and the risk of death from running a marathon. The main justification for using the Micro-Mort is that it represents a time tested method that is used by the US and UK Departments of Transport for calculating the value of a statistical life. Although the calculations by the US Department of Transport and the UK Department of Transport are different, both departments are using the MicroMort to calculate the 'Willingness to Pay' to estimate the value of preventing fatality [56].

### 6. Applying the proposed model for IoT MicroMort calculations

To test, validate and verify the findings of the new model, (a) the IoTMM for 2017 is calculated; and (b) the IoTMM for 2020 is forecasted, from using the following available data. In 2017, there were an estimated 378 million devices potentially vulnerable to hacking out of 8.4 billion connected IoT devices (or things) [57]. These numbers emerged from the BullGuard's IoT Scanner, where 310,000 users scanned their networks for vulnerabilities and 4.5 percent (nearly 14,000 devices), were reported as 'could be easily hacked', and from the Garner report that 8.4 billion connected things would be in use worldwide in 2017 [58]. To forecast the IoTMM for 2020, we use the forecasted data in the same report that predicts that the number of IoT connected devices will reach 20.4 billion by 2020, with more than 900 million potentially vulnerable devices by that date.

Therefore,

(a) the IoTMM for 2017 is calculated as 0.045 and
(b) the IoTMM for 2020 is calculated as 0.044

The next step is to calculate the enterprises' 'willingness to pay' to reduce 1 IoTMM for a specific category of devices. This is representative of the cost sum for an enterprise to accept a one-in-a-million IoTMM, or the cost sum that enterprise might be willing to pay to avoid a one-in-a-million chance of IoTMM. For the purposes of testing this model, we could apply a nominal Value of a Statistical Life (VSL) or the Value for Preventing a Fatality (VPF) [56] to evaluate the cost-effectiveness of expenditure on cyber security. IoT security spending is estimated to increase to $840.5 million in 2020 [59]. This can be used to define the IoT market value of 1 IoTMM in 2020 as $840.5 million divided by 20.4 billion or the number of IoT connected devices in 2020 [58]. However, it is important to understand what the value of 1 IoTMM represents in this scenario. We can explain this with an example, e.g. each T in a sample of 100,000 T's, represents the enterprise willingness to pay for a reduction in their individual IoT risk of 1 in 100,000, or 0.001%, over the next year. This reduction in risk would mean that we expect one fewer IoTMM among the sample of 100,000 $T_h$'s over the next year on average. Supposing that the answer was $0.00412, then the total dollar amount that the group would be willing to pay to save one statistical life in a year would be $0.00412 per T × 100,000 T's, or $4120. This is a very generic estimate that cannot be used by governments as guidance point for creating standards and legislation. Instead, the example describes the process of applying the formula, with a specific data sets for specific IoT devices. Governments and organisations would need to collect such data and be confident in the validity of the data before they can use the calculations for guidance points. In other words, the objective of this example is not to present a specific numbers for policy guidance, but to present a new formula that is multiple to different calculations, that can be applied for estimates when designing new policies, standards and legislations.

Calculating the IoTMM for 8.4 billion connected things would result in a number far greater than the estimated IoT security spending of $840.5 million in 2020. Unfortunately, we have no data on how the experts estimated the IoT security spending and the utility functions in such estimates are often not linear. Therefore, the economic value of 1 IoTMM does not represent a precise calculation of the value and risk. It represents more of a guidance point to show that, as more IoT devices become connected, their cyber security is not competitively priced, which increases the risk, and we need to be aware that we have no precise calculation of the IoT cyber risk or ICT cyber risk in general.

Enterprises can obtain a more precise valuation to their $T_h$'s by assessing the previously described valuation formula where $T_h$'s digital asset inventory D equals {D1, D2, . . ., Dn}; combined with the value of each asset V = {V1, V2, . . ., Vn}; and fDi represents the amount of residual risk Di is exposed to, measured in IoTMMD is. Resulting with the calculation of the value of 1 IoTMMD in 2020 as the amount of money $T_h$ is willing to pay to reduce 1 IoTMMD for its class D assets, valued with:

$$V = \sum_{i=1}^{Nc} CVi + \sum_{j=1}^{No} OVj$$

### 7. Analysis of results

The calculations and examples applied in the previous section represent real data that was applied to verify that the MM can be applied to assess the economic impact of IoT cyber risk, as described in the new model. Since there is no International IoT Asset Classification (IIoTAC) and no established Key IoT Cyber Risk Factors (KIoTCRF), the calculations of the new model serve just to verify the new model. After the establishment of IIoTAC and KIoTCRF, the new model could be applied to calculate more precise 'willingness to pay' that $T_h$ is willing to pay to reduce 1 IoTMMD.

We need to mention that the local linearity of the utility curve means that the MicroMort is useful for small incremental risks and rewards, not necessarily for large risks. Therefore, the IoTMM is not an ideal measure to calculate the IoT risk. Instead, IoTMM is better placed to measure for a given $T_h$ willingness to pay to reduce 1 IoTMMD for its class D assets.

Finally, we need to discuss the lack of IoT data. For example, the latest forecast from Gartner Inc. predicted that worldwide information security spending would reach $86.4 billion in 2017 and $93 billion in 2018. That forecast does not cover the IoT, ICS (Industrial Control Systems) and IIoT (Industrial Internet of Things) security [60]. Given the lack of data on IoT cyber risk, cyber loss or profits from different IoT vectors, it is extremely difficult to conduct IoT cyber risk analysis and argue on the soundness of the analysis. Since cyber insurance is in its infancy, insurance companies have not mastered the valuation of cyber risk in general. For example, Target was insured for $100 million of cyber risk in 2017, and suffered over $450 million of loss, with an estimated total of $1 billion by the end of 2017 [61]. This example clearly states that cyber insurance needs a lot more data to calculate, correlate and transfer risk with an acceptable degree of certainty. While general cyber risk cannot be calculated, the emergence of IoT has created new IoT risk verticals that are not at all defined in the cyber insurance policies.

### 7.1. Case study examples of IoT MicroMort calculations

The following calculations are conducted with secondary data from the The Hunt for IoT: The Rise of Thingbots [62] and the Gartner report [58] among other online reports.

1. The Persirai Thingbot is a Mirai code variant that scans for growth techniques and has affected at least 1250 IP camera models. The number of Persirai-infected hosts (IP-based cameras) in 2017 was 600,000. Therefore, the IoT MicroMort for IP-based cameras being infected by the Persirai Thingbot is 0.0714.
2. The Mirai Thingbot is a Distributed Denial of Service (DDoS) attack, composed of DVRs, routers, CCTV cameras. The Mirai source code was made public and is and adapted in other malware projects [62]. At present, there are Vigilante Thingbots like Halime or BrickerBot, that have been launched post-Mirai to destroy IoT devices before they can be infected and weaponised. The Vigilante Thingbots are grey hat efforts. In January 2017, the Vigilante Thingbots destroyed 18 million IoT devices. Therefore, the IoT MicroMort for IoT device being destroyed by a Vigilante Thingbot in January 2017 was 0.0021.

These examples are only for illustrative purposes and aim to describe the simplicity of calculating the IoT MicroMort with real case study data. One obvious weakness is present in these calculations. That is, the number of affected IoT devices is specific to the specification of the Mirai code for the Persirai Thingbot and the Vigilante Thingbots. The number of IoT devices that can be affected is limited to IP-based cameras for Persirai and DVRs, routers, and CCTV cameras for Mirai. The MicroMort values were calculated with the total number of IoT devices from the Garner report [58]. This is an obvious flaw of the calculation, but not of the model. Malware authors face the same problem as they have to write a custom package for a specific device. Frameworks are appearing in the malware market designed to enable broad ability to infect many different kinds of OS. They operate in the method where the operator plugs in a malware weapons package (goal is to destroy, or to subvert into a botnet or to compromise for some other purpose) and the framework provides a much expanded target population. These framework solution also helps the micromort problem, because with these frameworks, now we can run an MM calculation on the framework. However, these calculations are only for illustrative purposes and, as the statistics on IoT attacks and total number of devices becomes available, these calculations would become more meaningful and allow having more confidence in their conclusions.

## 8. Discussion

The objective of this study was to adapt the Cyber Value at Risk model and the MicroMort, established models for predicting uncertainty, for calculating the economic impact of IoT cyber risks. This resulted in a new model for calculating IoT risk called IoT MicroMort based on the Value at Risk model, which represents a tried and tested model for measuring the maximum possible loss over a given time period.

Our research study distinguished between IoT risk vectors and vertices (a concept that this article also proposed) and provided detailed explanation on the differences between the terminology. This re-categorisation of terminology is necessary as the concept of IoT is expanding at a fast rate and it is concerning that at this stage of the IoT evolution, we still do not have clear IoT terminology [63]. Our literature review expanded upon these new categories and proposed a taxonomic classification of the risk assessment requirements.

A Strengths, Weaknesses, Opportunities and Threats (SWOT) analysis was conducted for cyber risk frameworks, methods, systems and models aimed at setting up the scene for a unified, global standardisation of impact assessment approach for IoT cyber risk. This analysis presented the initial aim of all frameworks for a common unified reference point and integrated risk assessment approach.

In the spirt of a unified approach, we started with a discussion of established models for costing cyber-crime. Then, a new perspective based on probability theory was presented to calculate the impact of IoT risk from vectors and vertices, using a categorisation specific to the current IoT landscape. Some examples include the argument that the probability of Y is sequentially dependent on the probability of T, where if the two probabilities are not independent of one another, the accuracy of Y somehow depends on the accuracy of T. These contributions enabled our study to adopt the Value at Risk for measuring market risk and the MicroMort to develop units of value for reducing the IoT risk.

The units we used are compliant with the International Digital Asset Classification (IDAC) and we proposed the development of an International IoT Asset Classification (IIoTAC). To enable this, our study proposed categorisation of IoT digital assets.

The resulting model was then tested and validated with real data from the BullGuard's IoT Scanner (over 310,000 scans) and the Garner report on IoT connected devices. Two calculations have been pursued, one for the current situation and one for the future forecasts of IoT installed devices and emerging vulnerabilities.

This model, to our knowledge, represents the first attempt to quantify the economic impact of IoT cyber risk. While the model is not all encompassing we believe the model could work well in combination with other IoT risk quantification developed in the future. There is no doubt that one quantification model cannot resolve all the uncertainty that is surrounding the IoT risk assessment and quantification. If we compare such analysis with other well established risk calculations, we could conclude that quantitative analysts use multiple models and formulas to conduct risk quantification. A quantitative analyst would not use only the NPV or ROI to create a financial report. Such quantitative reports apply various models and formulas (e.g. Discounted Cash Flow, Time Value of Money) to create further report. It is indeed difficult to imagine that the economic aspects of cyber security can be assessed without referring to the cost of precise technical solutions (ROI for instance). However, the IoT MicroMort model proposed could serve as one model, from a range of new models, that could be used to calculate the IoT cyber risk in the future.

It is also worth mentioning that the proposed model has inherited some of the natural weaknesses of the MicroMort and Cyber Value at Risk models. It has taken decades to confirm the validity and the value of the MicroMort model. Today, the MicroMort is used to calculate many different types of risk occurrence. We are glad that we could add the IoT risk to the vast types of risk calculations conducted by the MicroMort.

## 9. Conclusion

The findings from this research lead to the conclusion that despite the many challenges in understanding the types and nature of cyber risk and their dependencies/interactions in the IoT space, there are acceptable ways to assess the economic impact of IoT and its risks. The article informs on how one may assess economic impact by making use of mathematical formalisms. The mathematical formalisms in the article are focused on IoT risk assessment CyVaR approach that aims to evaluate the future, rather than explain the past by means of historical analysis.

The multiple complexities explained in the study in terms of calculating the economic impact of IoT cyber risk lead to the conclusion that impact can only be assessed with (1) new risk metrics, and (2) a new valuation method specific for the new risk metrics, combined with (3) new regulatory framework and standardisation IoT databases, with (4) new risk vectors as defined in the form of International IoT Asset Classification (IIoTAC) and Key IoT Cyber Risk Factors (KIoTCRF).

This article presents new risk metrics, by adapting established methods for calculating risks and uncertainties, and identifies specific challenges for calculating the economic impact of IoT cyber risk (because of the vast number of different IoT vectors and vertices). The article further proposes a taxonomic classification of cyber risk assessment requirements (identification, estimation and prioritisation) and combines common basic terminology, common approaches and incorporates existing standards into a new model for calculating the economic impact of IoT cyber risk. The new risk metrics enable measuring the IoT risk, while the risk model enables the establishment of an acceptable IoT risk level. The adapted CyVaR determines the maximum loss sensitivity and enables adjusting the acceptable IoT risk level, by calculating the risk metrics from new operating conditions.

The new IoT MicroMort model integrates established risk assessment approaches with existing cyber risk assessment frameworks to develop a quantification tool for calculating the economic impact of IoT cyber risk.

### 9.1. Limitations

The IoT MicroMort model presented in this article represents a quantitative description of the overall IoT MicroMort risk. The article describes the model and how the model can be adapted to quantify specific IoT risks in individual sectors and products. The quantification does not represent the specific IoT risk for individual sectors and products. Rather, individual quantification and specific data for the model are required in order to present and reflect specific IoT risk for individual sectors and products. Nevertheless, the article presents the premises for a unified and integrated risk assessment approach and the IoT MicroMort model proposed could serve as one valid model to calculate the IoT cyber risk in the future.

The numbers presented in the article serve to verify the new model quantitatively. In addition, the value of the model in assessing costs of incidents and thus the economic risk should be considered in a similar way to the traditional application of the MicroMort model. The IoT MicroMort follows the same logic. That is, the IoT MicroMort calculates a probability for a certain event to occur, rather than an absolute value that describes the even. But this is the nature of probability itself.

In addition, the lack of IIoTAC and KIoTCRF prevent verification/validation of the new model for an absolute certainty of any MicroMort probability. Possibly the greatest challenge in valuing cyber risk resides in the unknown unknowns, i.e. the potential vulnerabilities in a Thing or ecosystem of Things that have yet to emerge and that can be exploitable by parties that are currently unidentified. This is a significant example of why MicroMort is as good of a methodology as we may be able to devise when quantifying the IoT risks.

**Funding sources**

This work was supported by the UK EPSRC with project [grant number EP/N02334X/1 and EP/N023013/1] and by the Cisco Research Centre [grant number 2017-169701 (3696)].

**Declarations of interest**

None.

**Acknowledgements**

Sincere gratitude to the Fulbright Commission.